\documentclass[12pt]{iopart}   

\usepackage{graphicx}

\begin{document}

\title[Subdiffusive random walk in a membrane system.]{Subdiffusive random walk in a membrane system. The generalized method of images approach}

\author{Tadeusz Koszto{\l}owicz} 
	
	\address{Institute of Physics, Jan Kochanowski University,
         ul. \'Swi\c{e}tokrzyska 15, 25-406 Kielce, Poland.}
         
	\eads{\mailto{tadeusz.kosztolowicz@ujk.edu.pl}}

\begin{abstract}
In this paper we study subdiffusion in a system with a thin membrane. At the beginning, the random walk of a particle is considered in a system with a discrete time and space variable and then the probability describing the evolution of the particle's position (Green's function) is transformed into a continuous system. Two models are considered differing here from each other regarding the assumptions about how the particle is stopped or reflected by the membrane when the particle attempts to pass through the membrane fails.
We show that for a system in which a membrane is partially permeable with respect to both its sides the Green's functions obtained for both models within the {\it continuous time random walk formalism} are equivalent to each other and expressed by the functions presented in the paper: T. Koszto{\l}owicz, Phys. Rev. E \textbf{91}, 022102 (2015), except the values defined at the membrane's surfaces. These functions generate the boundary condition at the membrane which contains a term with the Riemann--Liouville fractional derivative which vanishes over sufficiently long time, this term is present even when normal diffusion occurs. This boundary condition can be interpreted as the particle's passing through the thin membrane being a `long memory' process for subdiffusion as well as for normal diffusion. We also show that for a system with a one--sidedly fully permeable membrane, the Green's functions obtained within both models are not equivalent to each other and generate a boundary condition which does not contain the term with a fractional time derivative.

We also present the generalized method of images which provides the Green's functions for the membrane system, obtained in this paper. This method, which has a simple physical interpretation, is of a general nature and, in our opinion, can be used to find the Green's functions for a system with a thin membrane in which various models of subdiffusion can be applied. As an example, we find the Green's functions for the particular case of a `slow--subdiffusion' process in a system with a thin membrane.

\end{abstract}

\submitto{}

\maketitle

\section{Introduction\label{sec1}}

Normal diffusion or subdiffusion in a membrane system widely occurs in biology and engineering sciences (see, for example, \cite{hobbie,luckey,hsieh}).
In this paper we consider subdiffusion in a one dimensional system with a thin membrane.

The most commonly used definition of subdiffusion is that this process is the random walk of particles in which 
\begin{equation}\label{eq0a}
\left\langle (\Delta x)^2\right\rangle=\frac{2D_\alpha t^\alpha}{\Gamma(1+\alpha)}\;,
\end{equation} 
where $\left\langle (\Delta x)^2\right\rangle$ is the mean square displacement of a particle, $\alpha$ is a subdiffusion parameter (subdiffusion exponent), $D_\alpha$ is a subdiffusion coefficient, and $\Gamma$ is the Gamma function. Subdiffusion occurs in media in which the particles' movement is strongly hindered due to the complex internal structure of the medium, such as, for example, in porous media or gels \cite{mk,kdm}. The subdiffusion is often described by the following subdiffusion equation with the Riemann--Liouville fractional time derivative \cite{mk} 
\begin{equation}\label{eq18a}
\frac{\partial P(x,t;x_0)}{\partial t}=D_\alpha\frac{\partial^{1-\alpha}}{\partial t^{1-\alpha}}\frac{\partial^2 P(x,t;x_0)}{\partial x^2}\;.
\end{equation}
The Riemann--Liouville fractional derivative is defined as being valid for $\vartheta>0$ (here $k$ is a natural number which fulfils $k-1\leq \vartheta <k$)
\begin{equation}\label{eq18b}
\frac{d^\vartheta f(t)}{dt^\vartheta}=\frac{1}{\Gamma(k-\vartheta)}\frac{d^k}{dt^k}\int_0^t{(t-t')^{k-\vartheta-1}f(t')dt'}\;.
\end{equation}
There arise a problem how to set boundary conditions at a thin membrane (see, for example, \cite{korabel,kpre,kdl,tk}).

The structure of Eq. (\ref{eq18b}) shows that the presence of the fractional derivative in a model can be interpreted as dealing with a `long--memory' process.
For $\alpha=1$, the process is usually identified as normal diffusion, which is considered as the Markov process. Normal diffusion can be interpreted as a particle's random walk in which the mean square displacement of a single jump length and the mean frequency of jumps are both finite. However, there are processes in which the anomalously long waiting time for a particle to take its next step is entangled with the anomalously large length of jumps in a special way that provides $\alpha=1$ \cite{dybiec}. As  concluded in \cite{dybiec} in order to define subdiffusion, relation (\ref{eq0a}) should be supplemented by an appropriate stochastic interpretation of the random walk process. Such a simple interpretation is given within the {\it continuous time random walk} (CTRW) model where the random walk is described by the probability density $\lambda(\rho)$ of a single jump length $\rho$ and a probability density $\omega(\tau)$ of time $\tau$, which is needed for the particle to take its next step. It is assumed that for normal diffusion both distributions have finite moments of all natural orders whereas for subdiffusion the mean value of $\omega(\tau)$ is infinite and the moments of $\lambda(\rho)$ are finite. In this paper we base our consideration on the random walk model on a lattice for which $\lambda(\rho)=\frac{1}{2}\delta(\rho-\epsilon)+\frac{1}{2}\delta(\rho+\epsilon)$ (at the vicinity of the membrane the definition of $\lambda(\rho)$ is slightly different), $\epsilon$ is the distance between discrete sites; in this paper $\delta$ denotes both the Kronecker symbol (for a discrete space variable) or the delta--Dirac function (for a continuous space variable). The kind of diffusion process is defined here by the function $\omega(\tau)$. Obtaining the results for a system with a discrete spatial variable, we pass into a continuous system in the limit $\epsilon \longrightarrow 0$, using the formulas presented in this paper.

In paper \cite{tk}, subdiffusion in a system with a thin stopping membrane was considered; the membrane was assumed to be partially permeable with respect to both its sides, $0<q_1,q_2\leq 1$. The Green's functions, obtained for this case, generate the boundary condition at the membrane which contains the term with the Riemann--Liouville fractional derivative; this term, which is present even when normal diffusion occurs, vanishes over sufficiently long time. This boundary condition can be interpreted as the particle's passing through the thin membrane being a `long memory' process. In this paper we extend the consideration to the case of a reflecting membrane, subdiffusion in a system with a one--sidedly fully permeable membrane is also studied within the models of reflecting or stopping membranes.

There are three main aims of this paper. Firstly, we make a comparison between the Green's functions obtained for the model in which the particle can be reflected from the membrane and the model in which the particle can be stopped by the membrane. We find the conditions under which these functions are not equivalent to each other. Secondly, we examine the boundary conditions at the thin membrane and check when these conditions do not contain `long memory' terms. At this point we consider in detail the case of a one--sidedly fully permeable membrane. Thirdly, we find the generalized method of images which provides the Green's functions for the membrane system derived in this paper within the CTRW model. Next, using the generalized method of images, we obtain the Green's functions for a system with a thin membrane, in which the special case of `slow subdiffusion' occurs. `Slow subdiffusion' is understood here as a random walk for which all fractional moments of $\omega(\tau)$ are infinite, $\left\langle \tau^\rho\right\rangle\equiv\int_0^\infty\tau^\rho \omega(\tau)d\tau=\infty$ for $\rho>0$ \cite{dk}.

The organization of this paper is as follows. In Section \ref{sec2}, we present the general assumptions of the method. As an example, the Green's function for a homogeneous system without a membrane is derived. In Section \ref{sec3} we determine the Green's functions for a system with a membrane which reflects particles with some probability. In Section \ref{sec4} we show the Green's functions for a system in which particles can be stopped by the membrane; the Green's functions for a system with membrane which is partially permeable with respect to both its sides are taken from paper \cite{tk}; the new result is obtained for the case of a one--sidedly fully permeable membrane. The generalized method of images is presented in Section \ref{sec5}. As an example of the usefulness of this method, we find the Green's function for `slow--subdiffusion' in a system with a thin membrane. A comparison of the Green's functions obtained from the models presented in the Sections \ref{sec3} and \ref{sec4} and a discussion of the new results obtained in this paper are presented in Section \ref{sec6}.

\section{The method\label{sec2}}

We focus our attention on deriving the probability density (Green's function) $P(x,t;x_0)$ of finding a particle at point $x$ after time $t$ under the condition that at the initial moment $t=0$, the particle was at point $x_0$ (here we consider a particle's random walk in a one--dimensional system). To find this function for a system with a thin membrane we use a particle's random walk model in a system in which the time $n$ and spatial variable $m$ are discrete; after this, we pass from discrete to continuous variables.

In a system with discrete variables, the particle's random walk is described by a difference equation whose general form reads
\begin{equation}\label{eq0b}
P_{n+1}(m;m_0)=\sum_{m'}p_{m,m'}P_n(m';m_0)\;,
\end{equation}
where $p_{m,m'}$ is the probability that a particle jumps from site $m'$ directly to site $m$$, P_n(m;m_0)$ is the probability of finding a particle at site $m$ after step $n$ and $m_0$ is the initial position of the particle
\begin{equation}\label{eq2}
P_0(m;m_0)=\delta_{m,m_0}\;.
\end{equation}
For normal diffusion and subdiffusion long jumps occur with a low probability, thus we assume that the particle can only jump to its adjacent position. It is not allowed to remain in the current occupied position in the particle after the time at which the jump should be executed unless the particle is stopped by the membrane.
Such a model is useful in modeling normal or anomalous diffusion processes in a system with a partially permeable thin membrane. The reason is that Eq. (\ref{eq0b}), applied to describe a random walk in a membrane system in which the homogeneity of the system is impaired at a single point, is solvable by means of the generating function method \cite{montroll65,montroll64,weiss}. 
After solving these equations (more precisely, after determining the generating function for the equations) one can make the transition from discrete to continuous variables. The transition from discrete to continuous time is performed as in the CTRW method while passing from a discrete to continuous space variable is performed using the formulas presented in this paper. The reason for the introduction of such formulas is the specific behavior of the particles in the vicinity of the partially permeable membrane. 

In \cite{chandrasekhar} the random walk in a homogeneous system  was considered assuming
\begin{eqnarray}\label{eq0c}
p_{m,m'}&=&\frac{1}{2}\delta_{m-1,m'}+\frac{1}{2}\delta_{m+1,m'}.
\end{eqnarray}
The considerations were also extended to the case of normal diffusion in a system with a fully reflecting or fully absorbing membrane.
In these systems, the `parity rule', meaning that the number of steps $n$ and the distance traveled by a particle $|m-m_0|$, are both even numbers or both odd numbers, is fulfilled.
For the system with a fully reflecting or fully absorbing membrane the Green's functions were obtained using  symmetry arguments. It was shown \cite{chandrasekhar} that if the wall were removed, for each particle's trajectory which passes the line representing the position of the wall, there exists a symmetrical trajectory with respect to the wall, which is solely located in the half--space bounded by the wall, in which the particle started its random walk. In this way it was shown that the Green's functions for these systems is the sum (for the system with a fully reflecting wall) or difference (for the system with a fully absorbing wall) of the Green's functions obtained for the homogeneous system without the wall, and the initial points $x_0$ of these functions are located symmetrically with respect to the wall. The Green's function can be interpreted as a normalized concentration profile of particles initially located at the initial point, which means the concentration of $N$ particles, divided by $N$, $N\gg 1$. By differently describing the method of images, the idea of this method is that the wall is replaced by the additional particles' source located in such a way that the particles' concentration generated by the particles' source located at $m_0$ and by the additional particles' source both fulfil the appropriate boundary condition at the wall: the vanishing of the particles' flux at the reflecting wall and the vanishing of the particles' concentration at the absorbing wall. The method of images appears to be useful tool in deriving Green's function for a system containing two walls.

The situation is different when a thin partially permeable membrane is located in the system.
The presence of the membrane can be taken into account in the model in two different ways which depend on assumptions about how a particle is stopped by the membrane. 
Assuming that a thin membrane is located between the $N$ and the $N+1$ site, there are two possibilities concerning the influence of the membrane on a particle's movement. In the first model, the particle can be `reflected' by the membrane, which means that after the jump the particle always changes position. If the particle's attempt to pass through the membrane from the $N$ to the $N+1$ site fails with probability $q_1$, the particle is reflected by the membrane from the $N$ to the $N-1$ site; if the particle tries to jump through the membrane from the $N+1$ to the $N$ site and this attempt fails with probability $q_2$, then the particle is reflected from the $N+1$ to the $N+2$ site. In this case we obtain (see Fig. 1).

\begin{figure}[!ht]
\centering
\includegraphics[height=3.0cm]{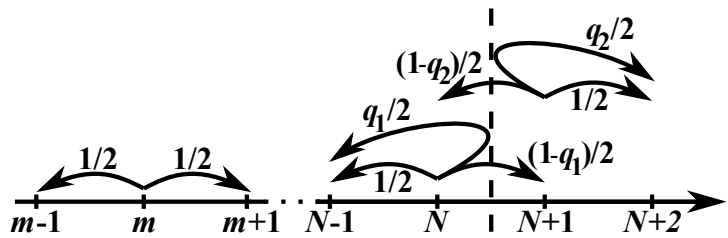}
\caption{System with reflecting membrane, more detailed description is in the text.}\label{Fig1}
\end{figure}

\begin{eqnarray}
p_{m,N}&=&\frac{1+q_1}{2}\delta_{N-1,m}+\frac{1-q_1}{2}\delta_{N+1,m}\;,\label{eq0e}\\
p_{m,N+1}&=&\frac{1-q_2}{2}\delta_{N,m}+\frac{1+q_2}{2}\delta_{N+2,m}\;,\label{eq0f}\\
p_{m,m'}&=&\frac{1}{2}\delta_{m-1,m'}+\frac{1}{2}\delta_{m+1,m'}\;,\;m'\neq N, N+1\label{eq0h}\;.
\end{eqnarray}
For this model the `parity rule' is satisfied.
  
For the model in which the particle can be stopped by the membrane, the particle remains in its position after the `jump' when the passing through the membrane has failed with probabilities $q_1$ and $q_2$. In this case we have (see Fig. 2).

\begin{figure}[!ht]
\centering
\includegraphics[height=3.0cm]{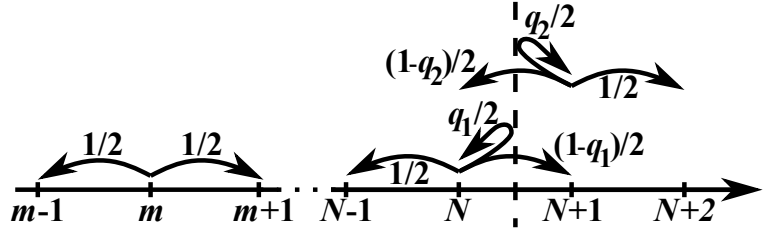}
\caption{System with stopping membrane, more detailed description is in the text.}\label{Fig2}
\end{figure}

\begin{eqnarray}
p_{m,N}&=&\frac{1}{2}\delta_{N-1,m}+\frac{q_1}{2}\delta_{N,m}+\frac{1-q_1}{2}\delta_{N+1,m}\;,\label{eq0i}\\
p_{m,N+1}&=&\frac{1}{2}\delta_{N+2,m}+\frac{q_2}{2}\delta_{N+1,m}+\frac{1-q_2}{2}\delta_{N,m}\;,\label{eq0j}\\
p_{m,m'}&=&\frac{1}{2}\delta_{m-1,m'}+\frac{1}{2}\delta_{m+1,m'}\;,\;m'\neq N, N+1\label{eq0k}\;.
\end{eqnarray}
For this case the `parity rule' is not satisfied. 
In the following, we will check if the assumptions involved in the models provide any noticeable differences between the Green's functions for a continuous system.

To solve the differential equations we use the generating function which is defined as
\begin{equation}\label{eq3}
  S(m,z;m_0)=\sum_{n=0}^{\infty}z^n P_n(m;m_0)\;.
\end{equation}
To pass from a discrete to continuous time we use the following formula
\begin{equation}\label{eq6}
  P(m,t;m_0)=\sum_{n=0}^{\infty}P_n(m;m_0)\Phi_n(t)\;,
\end{equation}
where $\Phi_n(t)$ is the probability that the particle takes $n$ jumps over time interval $(0,t)$.
This function fulfils the following relation \cite{mk} $\Phi_n(t)=\int_0^t U(t-t')Q_n(t')dt'$, where $U(t)=1-\int_0^t \omega(t')dt'$ is the probability that the particle does not perform any jump over time interval $(0,t)$, $\omega(t)$ is the probability distribution of the time which is needed for the particle to take its next step, $Q_n(t)$ denotes the probability that the particle performs $n$ steps over this time interval and the last step is taken exactly at time $t$, $Q_n(t)=\int_0^t Q_{n-1}(t-t')\omega(t')dt'$, $n>1$, $Q_1(t)=\omega(t)$, $Q_0(t)=\delta(t)$. 
In terms of the Laplace transform, $\mathcal{L}[f(t)]\equiv \hat{f}(s)=\int_0^\infty{{\rm e}^{-st}f(t)dt}$, the function $\Phi_{n}(t)$ reads \cite{mk}
\begin{equation}\label{eq7}
  \hat{\Phi}_n(s)=\frac{1-\hat{\omega}(s)}{s}\left[\hat{\omega}(s)\right]^n\;.
\end{equation}
Combining the Laplace transform of Eq. (\ref{eq6}) with Eqs. (\ref{eq3}) and (\ref{eq7}) we obtain
\begin{equation}\label{eq8} 
\hat{P}(m,s;m_0)=\frac{1-\hat{\omega}(s)}{s}S\left(m,\hat{\omega}(s);m_0\right)\;.
\end{equation}
To pass from a discrete to continuous spatial variable we
suppose 
\begin{equation}\label{eq11}
x=\epsilon m\;,\; x_0=\epsilon m_0\;,\;x_N=\epsilon N\;,
\end{equation} 
where $\epsilon$ denotes the distance between discrete sites. 
To pass form probability $\hat{P}(m,s;m_0)$ to spatial probability denisty $\hat{P}(x,s;x_0)$ we use the following relation valid for small values of $\epsilon$
\begin{equation}\label{eq12}
P(m,t;m_0)= \epsilon P(x,t;x_0)\;,
\end{equation}
and finally we take limit $\epsilon\longrightarrow 0$. Further considerations are performed assuming that $s$ is small, which corresponds to the case of large time due to Tauberian theorems. 
In practice, the limit of small $s$ means that only the leading terms with respect to this variable will be present in the Laplace transform of the Green's function whereas the limit of small $\epsilon$ means that this parameter will be absent in this function.

As an example we consider subdiffusion in a homogeneous system without bias. Starting from the following differential equation
\begin{equation} \label{eq1}
P_{n+1}(m;m_0)=\frac{1}{2}P_{n}(m-1;m_0)+\frac{1}{2}P_{n}(m+1;m_0),
\end{equation}
from Eqs.  (\ref{eq2}), (\ref{eq3}) and (\ref{eq1}) we get (see \cite{montroll65,barber} and the Appendix in this paper)
\begin{equation}\label{eq4}
S(m,z;m_0)=\frac{\eta^{|m-m_0|}(z)}{\sqrt{1-z^2}}\;,
\end{equation}
where
\begin{equation}\label{eq5}
\eta(z)=\frac{1-\sqrt{1-z^2}}{z}\;.
\end{equation}
In the standard {\it continuous time random walk} model it is assumed that $\hat{\omega}(s)$ for small $s$ reads \cite{mk}
\begin{equation}\label{eq9}
\hat{\omega}(s)= 1-\tau_\alpha s^\alpha\;,
\end{equation}
where $\tau_\alpha$ is a parameter which, together with $\alpha$, fully characterizes time distribution $\omega(t)$.
Eqs. (\ref{eq5}) and (\ref{eq9}) provide, for small values of $s$
\begin{equation}\label{eq10}
\eta(\hat{\omega}(s))= 1-\sqrt{2\tau_\alpha s^\alpha}\;.
\end{equation}
The subdiffusion coefficient is defined to be
\begin{equation}\label{eq13}
	D_\alpha=\frac{\epsilon^2}{2\tau_\alpha}\;.
\end{equation}
From Eqs. (\ref{eq8})--(\ref{eq13}) we observe
\begin{equation}\label{eq14}
\hat{P}(x,s;x_0)=\frac{1}{2\sqrt{D_\alpha}s^{1-\alpha/2}}\left(1-\epsilon\frac{s^{\alpha/2}}{\sqrt{D_\alpha}}\right)^{\frac{|x-x_0|}{\epsilon}}\;.
\end{equation}
Formally, the transition to a continuous variable was obtained by calculating limit $\epsilon\longrightarrow 0$. However, due to Eq. (\ref{eq13}) this provides $\tau_\alpha\longrightarrow 0$ which means that $\omega(t)$ is beyond any physical interpretation. In order to avoid problems of interpretation, we assume that $\epsilon$ is finite but small enough so that the following relation is satisfied (see also the discussion presented in \cite{tk})
\begin{equation}\label{eq15}
\left(1-\epsilon\frac{s^{\alpha/2}}{\sqrt{D_\alpha}}\right)^{\frac{|x-x_0|}{\epsilon}}\approx {\rm e}^{-\frac{|x-x_0|s^{\alpha/2}}{\sqrt{D_\alpha}}}\;.
\end{equation}
From Eqs. (\ref{eq14}) and (\ref{eq15}) we have
\begin{equation}\label{eq16}
\hat{P}(x,s;x_0)=\frac{1}{2\sqrt{D_\alpha}s^{1-\alpha/2}}{\rm e}^{-\frac{|x-x_0|s^{\alpha/2}}{\sqrt{D_\alpha}}}\;.
\end{equation}
The inverse Laplace transform of Eq. (\ref{eq15}), calculated by means of the following formula \cite{koszt}
\begin{equation}\label{eq17}
\mathcal{L}^{-1}\left[s^\nu {\rm e}^{-as^\beta}\right]\equiv f_{\nu,\beta}(t;a)
=\frac{1}{t^{\nu+1}}\sum_{k=0}^\infty{\frac{1}{k!\Gamma(-k\beta-\nu)}\left(-\frac{a}{t^\beta}\right)^k}\;,
\end{equation}
$a,\beta>0$, reads
\begin{equation}\label{eq18}
P(x,t;x_0)=\frac{1}{2\sqrt{D_\alpha}}\;f_{\alpha/2-1,\alpha/2}\left(t;\frac{|x-x_0|}{D_\alpha}\right)\;.
\end{equation}
We add that the function $f_{\nu,\beta}$ can be also expressed in terms of the Fox function.
Function (\ref{eq18}) is the solution of a fractional subdiffusion equation with a Riemann--Liouville fractional derivative (\ref{eq18a}) for the initial condition $P(x,0;x_0)=\delta(x-x_0)$.

\section{Random walk model with particle reflection from a membrane \label{sec3}}

Let us suppose that the thin membrane is located between the $N$ and the $N+1$ site.
We assume that the particle can be reflected from the membrane with probability $q_1$ when trying to jump from the $N$ to the $N+1$ site and with probability $q_2$ when trying to jump from the $N+1$ to the $N$ site (see Fig. \ref{Fig1}). Combining Eq. (\ref{eq0b}) and Eqs. (\ref{eq0e})--(\ref{eq0h}) we can observe
\begin{eqnarray}
 \label{eq19}P_{n+1}(N-1;m_0)&=&\frac{1}{2}P_{n}(N-2;m_0)+\frac{1+q_1}{2}P_{n}(N;m_0),\\ 
      \nonumber\\
 \label{eq20}P_{n+1}(N;m_0)&=&\frac{1}{2}P_{n}(N-1;m_0)+\frac{1-q_2}{2}P_{n}(N+1;m_0),\\ 
      \nonumber\\
 \label{eq21}P_{n+1}(N+1;m_0)&=&\frac{1-q_1}{2}P_{n}(N;m_0)+\frac{1}{2}P_{n}(N+2;m_0),\\ 
      \nonumber\\
 \label{eq22}P_{n+1}(N+2;m_0)&=&\frac{1+q_2}{2}P_{n}(N+1;m_0)+\frac{1}{2}P_{n}(N+3;m_0),\\
      \nonumber\\
 \label{eq23}P_{n+1}(m;m_0)&=&\frac{1}{2}P_{n}(m-1;m_0)+\frac{1}{2}P_{n}(m+1;m_0),\\
& &    m\neq N-1, N, N+1, N+2,\nonumber
\end{eqnarray}
the initial condition is given by Eq. (\ref{eq2}).

In the following, the functions $S$ and $P$ will be labelled by the indexes $ij$, which denote the signs of $m-N$ and $m_0-N$, respectively. We consider the case of $m_0\leq N$.
From Eqs. (\ref{eq2}), (\ref{eq3}) and (\ref{eq19})--(\ref{eq23}) we obtain (the details of the calculation are presented in the Appendix) 
\begin{eqnarray}
S_{--}(m,z;m_0)&=&\frac{\eta^{|m-m_0|}}{\sqrt{1-z^2}}+K_1(z)\frac{\eta^{2N-m-m_0}}{\sqrt{1-z^2}}\;,\;m\leq N-1,\label{eq24}\\
   \nonumber\\
S_{--}(N,z;m_0)&=&K_{1N}(z)\frac{\eta^{N-m_0}}{\sqrt{1-z^2}}\;,\label{eq25}\\
   \nonumber\\
S_{+-}(m,z;m_0)&=&K_2(z)\frac{\eta^{m-m_0}}{\sqrt{1-z^2}}\;,\;m\geq N+2,\label{eq27}\\
   \nonumber\\
S_{+-}(N+1,z;m_0)&=&K_{2N}(z)\frac{\eta^{N+1-m_0}}{\sqrt{1-z^2}}\;,\label{eq26}
\end{eqnarray}
where
\begin{eqnarray}
K_1(z)=\frac{q_1-q_2\eta^2(z)}{1-q_1 q_2\eta^2(z)}\;,\;
K_{1N}(z)=\frac{1-q_2\eta^2(z)}{1-q_1 q_2\eta^2(z)}\;,\label{eq28}\\
   \nonumber\\
K_2(z)=\frac{(1-q_1)(1+q_2)}{1-q_1 q_2\eta^2(z)}\;,\;
K_{2N}(z)=\frac{1-q_1}{1-q_1 q_2\eta^2(z)}\;.\label{eq29}
\end{eqnarray}  

Functions (\ref{eq28}) and (\ref{eq29}) play a key role in the Green's function for the membrane system, because only these functions depend on the permeability coefficients of the membrane. The main difficulty of finding the Green's functions for the continuous system is to find a suitable form of functions (\ref{eq28}) and (\ref{eq29}) for $\epsilon\longrightarrow 0$. We assume that in this limit, the Green's function must depend on the parameters of membrane permeability.
As we will see below, the case of $q_1,q_2>0$ and the case of $q_1=0$ or $q_2=0$ should be considered separately.

Similarly to \cite{tk}, the probabilities characterizing membrane permeability are assumed to be functions of $\epsilon$; this function for small values of argument $\epsilon$ reads
\begin{eqnarray}
q_1(\epsilon)=1-\frac{\epsilon^\sigma}{\gamma_1}\;,\label{eq30}\\
q_2(\epsilon)=1-\frac{\epsilon^\sigma}{\gamma_2}\;,\label{eq30a}
\end{eqnarray}
$\gamma_1$ and $\gamma_2$ being the membrane permeability coefficients defined for the continuous system, and $\sigma$ being the parameter to be determined. The reason for the introduction of these equations is that the frequency of jumps performed by the particle increases to infinity when $\epsilon$ drops to zero. For very small value of $\epsilon$, the frequency takes an `anomalously' large value \cite{tk}. 
A very large number of attempts to pass through the partially permeable membrane made over an arbitrarily small time interval means that the particle passes trough the membrane with probability equals to one; then, the membrane loses its selective property.

\subsection{The case $0<q_1\leq 1$ and $0<q_2\leq 1$ \label{sec3.1}}

From Eqs. (\ref{eq10}), (\ref{eq13}), and (\ref{eq28})--(\ref{eq30a}) we obtain
\begin{eqnarray}
\lefteqn{K_1(\hat{\omega}(s))=\frac{\epsilon^{\sigma-1}\left(\frac{1}{\gamma_1}-\frac{1}{\gamma_2}\right)+2\sqrt{\frac{s^{\alpha/2}}{D_\alpha}}}
{\epsilon^{\sigma-1}\left(\frac{1}{\gamma_1}+\frac{1}{\gamma_2}\right)+2\sqrt{\frac{s^{\alpha/2}}{D_\alpha}}}\;,}\label{eq31}\\
\lefteqn{K_{1N}(\hat{\omega}(s))=\frac{\epsilon^{\sigma-1}\frac{1}{\gamma_2}+2\sqrt{\frac{s^{\alpha/2}}{D_\alpha}}}
{\epsilon^{\sigma-1}\left(\frac{1}{\gamma_1}+\frac{1}{\gamma_2}\right)+2\sqrt{\frac{s^{\alpha/2}}{D_\alpha}}}\;,}\label{eq32}\\
   \nonumber\\
\lefteqn{K_2(\hat{\omega}(s))=\frac{2\epsilon^{\sigma-1}\frac{1}{\gamma_1}}
{\epsilon^{\sigma-1}\left(\frac{1}{\gamma_1}+\frac{1}{\gamma_2}\right)+2\sqrt{\frac{s^{\alpha/2}}{D_\alpha}}}\;,}\label{eq33}\\
\lefteqn{K_{2N}(\hat{\omega}(s))=\frac{\epsilon^{\sigma-1}\frac{1}{\gamma_1}}
{\epsilon^{\sigma-1}\left(\frac{1}{\gamma_1}+\frac{1}{\gamma_2}\right)+2\sqrt{\frac{s^{\alpha/2}}{D_\alpha}}}\;.}\label{eq34}
\end{eqnarray} 

The only possibility that Green's functions depend on the parameters of membrane permeability in the limit of small $\epsilon$, also for the case of symmetric membrane for which $\gamma_1=\gamma_2$, is $\sigma=1$. Then, form Eqs. (\ref{eq31})--(\ref{eq34}) we get for small $s$
\begin{eqnarray}
\lefteqn{K_1(\hat{\omega}(s))=\frac{\gamma_1-\gamma_2}{\gamma_1+\gamma_2}+\frac{4\gamma_2 \gamma_w s^{\alpha/2}}{(\gamma_1+\gamma_2)\sqrt{D_\alpha}}\;,}\label{eq35}\\
K_{1N}(\hat{\omega}(s))=\frac{\gamma_1}{\gamma_1+\gamma_2}+\frac{2\gamma_2 \gamma_w s^{\alpha/2}}{(\gamma_1+\gamma_2)\sqrt{D_\alpha}}\;,\label{eq36}\\
   \nonumber\\
K_2(\hat{\omega}(s))=\frac{2\gamma_2}{\gamma_1+\gamma_2}-\frac{4\gamma_2 \gamma_w s^{\alpha/2}}{(\gamma_1+\gamma_2)\sqrt{D_\alpha}}\;,\label{eq38}\\
K_{2N}(\hat{\omega}(s))=\frac{\gamma_2}{\gamma_1+\gamma_2}-\frac{2\gamma_2 \gamma_w s^{\alpha/2}}{(\gamma_1+\gamma_2)\sqrt{D_\alpha}}\;,\label{eq37}
\end{eqnarray}  
where $\gamma_w=\frac{\gamma_1 \gamma_2}{\gamma_1 +\gamma_2}$.

From Eqs. (\ref{eq8})--(\ref{eq12}), (\ref{eq24})--(\ref{eq26}) and (\ref{eq35})--(\ref{eq37}) we can observe
\begin{eqnarray}\label{eq39}
\hat{P}_{--}(x,s;x_0)=\frac{1}{2\sqrt{D_\alpha}s^{1-\alpha/2}}\left[{\rm e}^{-\frac{|x-x_0|s^{\alpha/2}}{\sqrt{D_\alpha}}}\right.\\
\left.+\left(\frac{\gamma_1-\gamma_2}{\gamma_1+\gamma_2}+\frac{4\gamma_2\gamma_w s^{\alpha/2}}{(\gamma_1+\gamma_2)\sqrt{D_\alpha}}\right){\rm e}^{-\frac{(2x_N-x-x_0)s^{\alpha/2}}{\sqrt{D_\alpha}}}\right]\;,\;x<x_N\nonumber
\end{eqnarray}
\begin{eqnarray}\label{eq40}
\hat{P}_{--}(x_N,s;x_0)=\frac{1}{2\sqrt{D_\alpha}s^{1-\alpha/2}}{\rm e}^{-\frac{(x_N-x_0)s^{\alpha/2}}{\sqrt{D_\alpha}}}\left(\frac{\gamma_1}{\gamma_1+\gamma_2}+\frac{2\gamma_2\gamma_w s^{\alpha/2}}{(\gamma_1+\gamma_2)\sqrt{D_\alpha}}\right)\;,
\end{eqnarray}
\begin{eqnarray}\label{eq43}
\hat{P}_{+-}(x,s;x_0)=\frac{1}{2\sqrt{D_\alpha}s^{1-\alpha/2}}\left[\frac{2\gamma_2}{\gamma_1+\gamma_2}-\frac{4\gamma_2 \gamma_w s^{\alpha/2}}{(\gamma_1+\gamma_2)\sqrt{D_\alpha}}\right]{\rm e}^{-\frac{(x-x_0)s^{\alpha/2}}{\sqrt{D_\alpha}}}\;,\\
\nonumber x>x_N\;,
\end{eqnarray}
\begin{eqnarray}\label{eq44}
\hat{P}_{+-}(x_N,s;x_0)=\frac{1}{2\sqrt{D_\alpha}s^{1-\alpha/2}}\left[\frac{\gamma_2}{\gamma_1+\gamma_2}-\frac{2\gamma_2 \gamma_w s^{\alpha/2}}{(\gamma_1+\gamma_2)\sqrt{D_\alpha}}\right]{\rm e}^{-\frac{(x_N-x_0)s^{\alpha/2}}{\sqrt{D_\alpha}}}\;.
\end{eqnarray}
The inverse Laplace transform of the above functions read
\begin{eqnarray}\label{eq41}
P_{--}(x,t;x_0)=\frac{1}{2\sqrt{D_\alpha}}\;f_{\alpha/2-1,\alpha/2}\left(t;\frac{|x-x_0|}{\sqrt{D_\alpha}}\right)\\
+\frac{\gamma_1-\gamma_2}{2\sqrt{D_\alpha}(\gamma_1+\gamma_2)}\;f_{\alpha/2-1,\alpha/2}\left(t;\frac{2x_N-x-x_0}{\sqrt{D_\alpha}}\right)\nonumber\\
+\frac{2\gamma_2 \gamma_w}{D_\alpha(\gamma_1+\gamma_2)}\;f_{\alpha-1,\alpha/2}\left(t;\frac{2x_N-x-x_0}{\sqrt{D_\alpha}}\right)\;,\;x<x_N\;,\nonumber
\end{eqnarray}
\begin{eqnarray}\label{eq42}
P_{--}(x_N,t;x_0)=\frac{\gamma_1}{2\sqrt{D_\alpha}(\gamma_1+\gamma_2)}\;f_{\alpha/2-1,\alpha/2}\left(t;\frac{x_N-x_0}{\sqrt{D_\alpha}}\right)\\
+\frac{\gamma_2 \gamma_w}{D_\alpha(\gamma_1+\gamma_2)}\;f_{\alpha-1,\alpha/2}\left(t;\frac{x_N-x_0}{\sqrt{D_\alpha}}\right)\;,\nonumber
\end{eqnarray}
\begin{eqnarray}\label{eq45}
P_{+-}(x,t;x_0)=\frac{\gamma_2}{\sqrt{D_\alpha}(\gamma_1+\gamma_2)}\;f_{\alpha/2-1,\alpha/2}\left(t;\frac{x-x_0}{\sqrt{D_\alpha}}\right)\\
-\frac{2\gamma_2 \gamma_w}{D_\alpha(\gamma_1+\gamma_2)}\;f_{\alpha-1,\alpha/2}\left(t;\frac{x-x_0}{\sqrt{D_\alpha}}\right)\;,\;x>x_N\;,\nonumber
\end{eqnarray}
\begin{eqnarray}\label{eq46}
P_{+-}(x_N,t;x_0)=\frac{\gamma_2}{2\sqrt{D_\alpha}(\gamma_1+\gamma_2)}\;f_{\alpha/2-1,\alpha/2}\left(t;\frac{x_N-x_0}{\sqrt{D_\alpha}}\right)\\
-\frac{\gamma_2 \gamma_w}{D_\alpha(\gamma_1+\gamma_2)}\;f_{\alpha-1,\alpha/2}\left(t;\frac{x_N-x_0}{\sqrt{D_\alpha}}\right)\;.\nonumber
\end{eqnarray}
It is easy to see that the Green's functions lose their continuity at  membrane surfaces
\begin{equation}\label{eq47}
P_{\pm -}(x_N,t;x_0)=\frac{1}{2}P_{\pm -}(x\longrightarrow x_N^\mp,t;x_0)\;.
\end{equation}

When the membrane is one--sidedly fully permeable, e.g. in the case of $q_1=0$ or $q_2=0$, the dependence of functions (\ref{eq28}) and (\ref{eq29}) on the parameter $\epsilon$ is different than in the previous case, which we will show in the next sections.

\subsection{The case $q_1=0$ and $0<q_2\leq 1$ \label{sec3.2}}

Supposing $q_1=0$ and $q_2>0$, combining Eqs. (\ref{eq10}), (\ref{eq13}), and (\ref{eq28}), (\ref{eq29}), and (\ref{eq30a}) we obtain over a limit of small values of $\epsilon$ and $s$
\begin{eqnarray}
K_1(\hat{\omega}(s))=-1+\frac{\epsilon^\sigma}{\gamma_2}+2\epsilon\sqrt{\frac{s^\alpha}{D_\alpha}}\;,\;
K_{1N}(\hat{\omega}(s))=\frac{\epsilon^\sigma}{\gamma_2}+2\epsilon\sqrt{\frac{s^\alpha}{D_\alpha}}\;,\label{eq48}\\
K_2(\hat{\omega}(s))=2-\frac{\epsilon^\sigma}{\gamma_2}\;,\; 
K_{2N}(\hat{\omega}(s))\equiv 1\;.\label{eq49}
\end{eqnarray} 
The Green's functions are dependent on parameter $\gamma_2$ over the limit of small values of $\epsilon$ only if $\sigma=0$; in this case, the functions are independent of  parameter $s$. 
Proceeding as in Sec. \ref{sec3.1} we obtain
\begin{eqnarray}\label{eq50}
P_{--}(x,t;x_0)=f_{\alpha/2-1,\alpha/2}\left(t;\frac{|x-x_0|}{\sqrt{D_\alpha}}\right)\\+\left(\frac{1}{\gamma_2}-1\right)f_{\alpha/2-1,\alpha/2}\left(t;\frac{2x_N-x-x_0}{\sqrt{D_\alpha}}\right)\;,\;x<x_N\;,\nonumber
\end{eqnarray}
\begin{eqnarray}\label{eq51}
P_{--}(x_N,t;x_0)=\frac{1}{\gamma_2} f_{\alpha/2-1,\alpha/2}\left(t;\frac{x_N-x_0}{\sqrt{D_\alpha}}\right)\;,
\end{eqnarray}
\begin{eqnarray}\label{eq53}
P_{+-}(x,t;x_0)=\left(2-\frac{1}{\gamma_2}\right) f_{\alpha/2-1,\alpha/2}\left(t;\frac{x-x_0}{\sqrt{D_\alpha}}\right)\;,\;x>x_N\;,
\end{eqnarray}
\begin{eqnarray}\label{eq52}
P_{+-}(x_N,t;x_0)=f_{\alpha/2-1,\alpha/2}\left(t;\frac{x_N-x_0}{\sqrt{D_\alpha}}\right)\;.
\end{eqnarray}

\subsection{The case $0<q_1\leq 1$ and $q_2=0$ \label{sec3.3}}

Supposing $q_2=0$ and $q_1>0$, from Eqs. (\ref{eq10}), (\ref{eq13}), (\ref{eq28}), (\ref{eq29}), and (\ref{eq30}) we obtain 
\begin{eqnarray}
K_1(\hat{\omega}(s))=1-\frac{\epsilon^\sigma}{\gamma_1}\;,\;
K_{1N}(\hat{\omega}(s))\equiv 1\label{eq54}\;,\\
K_2(\hat{\omega}(s))\equiv K_{2N}(\hat{\omega}(s))=\frac{\epsilon^\sigma}{\gamma_1}\;.\label{eq55}
\end{eqnarray} 
Thus, Eqs. (\ref{eq54}) and (\ref{eq55}) depend on the membrane permeability coefficient over the limit of small values of $\epsilon$ only if $\sigma=0$.
After calculations, we have the following Green's functions

\begin{eqnarray}\label{eq56}
P_{--}(x,t;x_0)=f_{\alpha/2-1,\alpha/2}\left(t;\frac{|x-x_0|}{\sqrt{D_\alpha}}\right)\\
+\left(1-\frac{1}{\gamma_1}\right) f_{\alpha/2-1,\alpha/2}\left(t;\frac{2x_N-x-x_0}{\sqrt{D_\alpha}}\right)\;,\;x<x_N\;,\nonumber
\end{eqnarray}
\begin{eqnarray}\label{eq57}
P_{--}(x_N,t;x_0)=f_{\alpha/2-1,\alpha/2}\left(t;\frac{x_N-x_0}{\sqrt{D_\alpha}}\right)\;,
\end{eqnarray}
\begin{eqnarray}\label{eq59}
P_{+-}(x,t;x_0)=\frac{1}{\gamma_1} f_{\alpha/2-1,\alpha/2}\left(t;\frac{x-x_0}{\sqrt{D_\alpha}}\right)\;,\;x>x_N\;,
\end{eqnarray}
\begin{eqnarray}\label{eq58}
P_{+-}(x_N,t;x_0)=\frac{1}{\gamma_1} f_{\alpha/2-1,\alpha/2}\left(t;\frac{x_N-x_0}{\sqrt{D_\alpha}}\right)\;.
\end{eqnarray}

\section{Random walk model with particle stopped by a membrane \label{sec4}}

In \cite{tk} we considered a particle's random walk model in a system with a thin membrane, in which the particle trying to pass the membrane from the $N$ to the $N+1$ site, may be stopped by the membrane with probability $q_1$ or pass through the membrane with probability $1-q_1$. The particle's halting by the membrane means that the particle does not change its position after its `jump'. When the molecule is trying to jump from position $N+1$ to $N$, the probability of the blocking of the particle through the membrane is $q_2$, and the probability of passage through the membrane equals $1-q_2$.

From Eqs. (\ref{eq0b}) and (\ref{eq0i})--(\ref{eq0k}) we have the following difference equations 
\begin{eqnarray}
 \label{eq60}P_{n+1}(m;m_0)&=&\frac{1}{2}P_{n}(m-1;m_0)+\frac{1}{2}P_{n}(m+1;m_0),\;
    m\neq N, N+1,\\
      \nonumber\\
  \label{eq61}P_{n+1}(N;m_0)&=&\frac{1}{2}P_{n}(N-1;m_0)+\frac{1-q_2}{2}P_{n}(N+1;m_0)\nonumber\\
  &+&\frac{q_1}{2}P_{n}(N;m_0),\\
      \nonumber\\
  \label{eq62}P_{n+1}(N+1;m_0)&=&\frac{1-q_1}{2}P_{n}(N;m_0)+\frac{1}{2}P_{n}(N+2;m_0)\nonumber\\
  &+&\frac{q_2}{2}P_{n}(N+1;m_0).
\end{eqnarray}
The generating functions for the equations (\ref{eq60})--(\ref{eq62}) read \cite{tk}
\begin{equation}\label{eq63}
S_{--}(m,z;m_0)=\frac{\eta^{|m-m_0|}(z)}{\sqrt{1-z^2}}
+\Lambda_1(z)\frac{\eta^{2N-m-m_0+1}(z)}{\sqrt{1-z^2}}\;,
\end{equation}
\begin{equation}\label{eq64}
S_{+-}(m,z;m_0)=\Lambda_2(z)\frac{\eta^{m-m_0}(z)}{\sqrt{1-z^2}}\;,
\end{equation}
where
\begin{eqnarray}
\Lambda_1(z)=\frac{q_1-q_2\eta(z)}{1-(q_1+q_2-1)\eta(z)}\;,\label{eq65}\\
\Lambda_2(z)=\frac{(1+\eta(z))(1-q_1)}{1-(q_1+q_2-1)\eta(z)}\label{eq66}\;.
\end{eqnarray}

\subsection{The case $0<q_1\leq 1$ and $0<q_2\leq 1$ \label{sec4.1}}

In the following $\tilde{\gamma}_1$ and $\tilde{\gamma}_2$ denote the permeability coefficients of the stopping membrane, defined for a system with a continuous spatial variable.
As shown in \cite{tk}, the probabilities of passing through the membrane should be chosen as the following functions of $\epsilon$
\begin{equation}\label{eq67}
q_1(\epsilon)=1-\frac{\epsilon}{\tilde{\gamma}_1}\;,\;q_2(\epsilon)= 1-\frac{\epsilon}{\tilde{\gamma}_2}\;,
\end{equation}
which, together with Eqs. (\ref{eq10}), (\ref{eq13}), (\ref{eq65}), (\ref{eq66}), and (\ref{eq67}) provide
\begin{eqnarray}
\lefteqn{\Lambda_1(\hat{\omega}(s))=\frac{\tilde{\gamma}_1-\tilde{\gamma}_2}{\tilde{\gamma}_1+\tilde{\gamma}_2}+\frac{2\tilde{\gamma}_2 \tilde{\gamma}_w s^{\alpha/2}}{(\tilde{\gamma}_1+\tilde{\gamma}_2)\sqrt{D_\alpha}}\;,}\label{eq65a}\\
\Lambda_2(\hat{\omega}(s))=\frac{2\tilde{\gamma}_2}{\tilde{\gamma}_1+\tilde{\gamma}_2}-\frac{2\tilde{\gamma}_2 \tilde{\gamma}_w s^{\alpha/2}}{(\tilde{\gamma}_1+\tilde{\gamma}_2)\sqrt{D_\alpha}}\;,\label{eq66a}
\end{eqnarray} 
where $\tilde{\gamma}_w=\frac{\tilde{\gamma}_1 \tilde{\gamma}_2}{\tilde{\gamma}_1 +\tilde{\gamma}_2}$. 
The Green's functions read \cite{tk}
\begin{eqnarray}\label{eq67a}
P_{--}(x,t;x_0)=\frac{1}{2\sqrt{D_\alpha}}\Bigg[f_{\alpha/2-1,\alpha/2}\left(t;\frac{|x-x_0|}{\sqrt{D_\alpha}}\right)\\
+\frac{\tilde{\gamma}_1-\tilde{\gamma}_2}{\tilde{\gamma}_1+\tilde{\gamma}_2} f_{\alpha/2-1,\alpha/2}\left(t;\frac{2x_N-x-x_0}{\sqrt{D_\alpha}}\right)\Bigg]\nonumber\\
+ \frac{\tilde{\gamma}_2 \tilde{\gamma}_w}{(\tilde{\gamma}_1+\tilde{\gamma}_2)D_\alpha}f_{\alpha-1,\alpha/2}\left(t;\frac{2x_N-x-x_0}{\sqrt{D_\alpha}}\right)\;,\nonumber
\end{eqnarray}
\begin{eqnarray}\label{eq67b}
P_{+-}(x,t;x_0)=\frac{2\tilde{\gamma}_2}{\tilde{\gamma}_1+\tilde{\gamma}_2}\frac{1}{2\sqrt{D_\alpha}}\;f_{\alpha/2-1,\alpha/2}\left(t;\frac{x-x_0}{\sqrt{D_\alpha}}\right)\\
- \frac{\tilde{\gamma}_2 \tilde{\gamma}_w}{(\tilde{\gamma}_1+\tilde{\gamma}_2)D_\alpha}\;f_{\alpha-1,\alpha/2}\left(t;\frac{x-x_0}{\sqrt{D_\alpha}}\right)\nonumber\;.
\end{eqnarray}

The Green's functions Eqs. (\ref{eq67a}) and (\ref{eq67b}) coincide with the Green's functions obtained for the system with a reflecting membrane, Eqs. (\ref{eq41}) and (\ref{eq45}), respectively, if $\gamma_1=\tilde{\gamma}_1/2$ and $\gamma_2=\tilde{\gamma}_2/2$.

As in Section \ref{sec3}, the case of a membrane which is one--sidedly fully permeable should be considered separately.

\subsection{The case $q_1=0$ and $0<q_2\leq 1$ \label{sec4.2}}

We suppose that $q_1=0$ and
\begin{equation}\label{eq67c}
q_2(\epsilon)= 1-\frac{\epsilon^\sigma}{\tilde{\gamma}_2}\;.
\end{equation}
Form Eqs. (\ref{eq10}), (\ref{eq13}), (\ref{eq65}), (\ref{eq66}) and (\ref{eq67c}) we obtain
\begin{equation}\label{eq68}
\Lambda_1(z)=\frac{1-\epsilon^\sigma\frac{1}{\tilde{\gamma}_2}-\epsilon\sqrt{\frac{s^\alpha}{D_\alpha}}}{1+\epsilon^\sigma\frac{1}{\tilde{\gamma}_2}}\;\;,\;\;
\Lambda_2(z)=\frac{2-\epsilon\sqrt{\frac{s^\alpha}{D_\alpha}}}{1+\epsilon^\sigma\frac{1}{\tilde{\gamma}_2}}\;.
\end{equation}
As previously, functions Eq. (\ref{eq68}) depend on the parameter of membrane permeability only for $\sigma=0$. In this case, the Green's functions read 
\begin{eqnarray}\label{eq69a}
P_{--}(x,t;x_0)=\frac{1}{2\sqrt{D_\alpha}}\Bigg[f_{\alpha/2-1,\alpha/2}\left(t;\frac{|x-x_0|}{\sqrt{D_\alpha}}\right)\\
+\frac{\tilde{\gamma}_2-1}{\tilde{\gamma}_2+1} f_{\alpha/2-1,\alpha/2}\left(t;\frac{2x_N-x-x_0}{\sqrt{D_\alpha}}\right)\Bigg]\nonumber\;,
\end{eqnarray}
\begin{eqnarray}\label{eq69b}
P_{+-}(x,t;x_0)=\frac{\tilde{\gamma}_2}{(\tilde{\gamma}_2 +1)\sqrt{D_\alpha}}\;f_{\alpha/2-1,\alpha/2}\left(t;\frac{x-x_0}{\sqrt{D_\alpha}}\right)\;.
\end{eqnarray}

\subsection{The case $0<q_1\leq 1$ and $q_2=0$ \label{sec4.3}}

Supposing $q_2=0$ and
\begin{equation}\label{eq67d}
q_1(\epsilon)=1-\frac{\epsilon^\sigma}{\tilde{\gamma}_1}\;,
\end{equation}
proceeding as in Sec. \ref{sec4.2} we obtain
\begin{equation}\label{eq69}
\Lambda_1(\hat{\omega}(s))=\frac{1-\epsilon^\sigma\frac{1}{\tilde{\gamma}_1}}{1+\epsilon^\sigma\frac{1}{\tilde{\gamma}_1}}\;\;,\;\;
\Lambda_2(\hat{\omega}(s))=\frac{2\epsilon^\sigma\frac{1}{\tilde{\gamma}_1}}{1+\epsilon^\sigma\frac{1}{\tilde{\gamma}_1}}\;.
\end{equation}
Similarly to the previous case, we have to take $\sigma=0$. The Green's functions read 
\begin{eqnarray}\label{eq69c}
P_{--}(x,t;x_0)=\frac{1}{2\sqrt{D_\alpha}}\Bigg[f_{\alpha/2-1,\alpha/2}\left(t;\frac{|x-x_0|}{\sqrt{D_\alpha}}\right)\\
+\frac{\tilde{\gamma}_1-1}{\tilde{\gamma}_1+1} f_{\alpha/2-1,\alpha/2}\left(t;\frac{2x_N-x-x_0}{\sqrt{D_\alpha}}\right)\Bigg]\nonumber\;,
\end{eqnarray}
\begin{eqnarray}\label{eq69d}
P_{+-}(x,t;x_0)=\frac{1}{(\tilde{\gamma}_1 +1)\sqrt{D_\alpha}}\;f_{\alpha/2-1,\alpha/2}\left(t;\frac{x-x_0}{\sqrt{D_\alpha}}\right)\;.
\end{eqnarray}

\section{Generalized method of images \label{sec5}}

As we briefly discussed in the Introduction, the classical version of the method of images has been used to determine the Green's function in a system containing a fully reflecting or fully absorbing wall. In this method, Green's function is considered to be the normalized concentration of diffusing particles beginning their movement from  initial point $x_0$ at $t=0$ .
The main idea of this method is to replace the wall by an additional source of particles in such a way that the concentration of particles generated by all particle sources occurring in the system, fulfils the boundary condition which is assumed at the wall. 
For a system with a fully reflecting wall, the boundary condition at the wall is that the diffusive flux vanishing at the wall, $J(x_N,t;x_0)=0$, $x_N$ is the position of the wall. The same effect would be achieved if the wall were to be replaced by an additional source of particles located symmetrically to point $x_0$ with respect to the wall. 
For a system with a fully absorbing wall, the boundary condition reads $P(x_N,t;x_0)=0$, the additional source located symmetrically to point $x_0$ with respect to the wall should be substracted from the Green's function representing the particles' source located at $x_0$.
Thus, assuming that $x_0<x_N$, the Green's function for the above mentioned cases can be written in the following compact form 
\begin{equation}\label{eq70}
P(x,t;x_0)=P_0(x,t;x_0)+\varsigma P_0(x,t;2x_N-x_0)\;, 
\end{equation}
where $\varsigma=1$ is for a fully reflecting wall and $\varsigma=-1$ is for a fully absorbing wall, $P_0$ denotes here the Green's function for homogeneous system with removed wall.

The form of the Green's functions, presented in Secs. \ref{sec3} and \ref{sec4}, shows that these functions can also be determined using the method of images which is understood here as a replacement of the membrane by the additional source function. Analyzing the structure of the Green's functions obtained in the previous section we note that the functions can be expressed by the following equations
\begin{equation}\label{eq71}
P_{--}(x,t;x_0)=P_0(x,t;x_0)+P_C(x,t;2x_N-x_0)\;,
\end{equation}
\begin{equation}\label{eq72}
P_{+-}(x,t;x_0)=P_0(x,t;x_0)-P_C(x,t;x_0)\;,
\end{equation}
where $P_C$ denotes a `compound' source function. This function has the following structure
\begin{equation}\label{eq73}
P_C(x,t;x_0)=\kappa_0 P_0(x,t;x_0)+\kappa_G P_G(x,t;x_0)\;,
\end{equation}
where
\begin{equation}\label{eq74}
P_G(x,t;x_0)=\left|\frac{d}{dx}P_0(x,t;x_0)\right|
\end{equation}
is a `gradient source function'. Parameters $\kappa_0$ and $\kappa_G$ depend on the membrane permeability coefficients. In general, $\kappa_0$ can be interpreted as the relative measure of the asymmetry of the membrane; $\kappa_0 =0$ for the symmetrical membrane. The particular cases are the following: for $0<q_1,q_2\leq 1$ for the system with a reflecting membrane we observe 
\begin{equation}\label{eq74a}
\kappa_0=\frac{\gamma_1-\gamma_2}{\gamma_1+\gamma_2}\;,\;\kappa_G=\frac{4\gamma_1}{(1+\gamma_1/\gamma_2)^2}\;.
\end{equation}
For the system with a stopping membrane, Eqs. (\ref{eq71})--(\ref{eq74a}) are still valid, with $\gamma_{1,2}=\tilde{\gamma}_{1,2}/2$. For the case of a fully reflecting wall we have $\gamma_1\longrightarrow\infty$, then $\kappa_0=1$ and $\kappa_G=0$ which means that a function $P_{--}$ takes the form of function (\ref{eq70}) with $\varsigma=1$ and $P_{+-}$ equal to zero. For a partially absorbing wall there is $\gamma_2\longrightarrow\infty$ and $0<\gamma_1<\infty$ which gives $\kappa_0=-1$ and $\kappa_G=4\gamma_1$. When the membrane is one--sidedly fully permeable, then $\kappa_G=0$ for all cases, but parameter $\kappa_0$ is not the same for both the models mentioned above. For the model of the reflecting membrane we have: for $q_1=0$, $q_2>0$ there is $\kappa_0=-1+1/\gamma_2$ and for $q_2=0$, $q_1>0$ there is $\kappa_0=1-1/\gamma_1$. For the model of the stopping membrane we have: for $q_1=0$, $q_2>0$ there is $\kappa_0=(-1+1/\tilde{\gamma}_2)/(1+1/\tilde{\gamma}_2)$ and for $q_2=0$, $q_1>0$ there is $\kappa_0=(1-1/\tilde{\gamma}_1)/(1+1/\tilde{\gamma}_1)$. 

We assume that the method of images may be used to determine the Green's function for various models of subdiffusion in a membrane system. As an example, we consider the `slow subdiffusion' model, in which the waiting time for the jump is given by the following function $\omega(t)=1-\left[\frac{{\rm ln}\mu}{{\rm ln}(\mu+t)}\right]^{r-1}$, $\mu,r>1$ \cite{dk}, which, for a long time, $t\gg \mu$, reads
\begin{equation}\label{eq75}
\omega(t)=1-\left(\frac{{\rm ln}\mu}{{\rm ln} t}\right)^{r-1}\;.
\end{equation}
Let the subdiffusion coefficient be defined 
\begin{equation}\label{eq75a}
D_{r,\mu}=\frac{\epsilon^2}{2({\rm ln}\mu)^{r-1}}\;,
\end{equation}
the Green's function for these processes for a homogeneous system without a membrane reads \cite{dk}
\begin{equation}\label{eq76}
P_0(x,t;x_0)=\frac{1}{2\sqrt{D_{r,\mu}({\rm ln}t)^{r-1}}}{\rm e}^{-\frac{|x-x_0|}{\sqrt{D_{r,\mu}({\rm ln}t)^{r-1}}}}\;.
\end{equation}
This Green's function provides the relation $\left\langle (\Delta x)^2\right\rangle=2D_{r,\mu}({\rm ln}t)^{r-1}$ \cite{dk}. Using Eq. (\ref{eq74}) we have
\begin{equation}\label{eq77}
P_G(x,t;x_0)=\frac{1}{2D_{r,\mu}({\rm ln}t)^{r-1}}{\rm e}^{-\frac{|x-x_0|}{\sqrt{D_{r,\mu}({\rm ln}t)^{r-1}}}}\;.
\end{equation}
The Green's functions for the slow subdiffusion process in a system with a thin membrane, derived by means of the generalized method of images, read
\begin{eqnarray}\label{eq77a}
P_{--}(x,t;x_0)=\frac{\kappa_0}{2\sqrt{D_{r,\mu}({\rm ln}t)^{r-1}}}{\rm e}^{-\frac{|x-x_0|}{\sqrt{D_{r,\mu}({\rm ln}t)^{r-1}}}}\\
+\left[\frac{\kappa_0}{2\sqrt{D_{r,\mu}({\rm ln}t)^{r-1}}}{\rm e}^{-\frac{(2x_N-x-x_0)}{\sqrt{D_{r,\mu}({\rm ln}t)^{r-1}}}}
+\frac{\kappa_G}{2D_{r,\mu}({\rm ln}t)^{r-1}}{\rm e}^{-\frac{(2x_N-x-x_0)}{\sqrt{D_{r,\mu}({\rm ln}t)^{r-1}}}}\right]\nonumber\;,
\end{eqnarray}
\begin{eqnarray}\label{eq77b}
P_{+-}(x,t;x_0)=\frac{1-\kappa_0}{2\sqrt{D_{r,\mu}({\rm ln}t)^{r-1}}}{\rm e}^{-\frac{(x-x_0)}{\sqrt{D_{r,\mu}({\rm ln}t)^{r-1}}}}\\
-\frac{\kappa_G}{2D_{r,\mu}({\rm ln}t)^{r-1}}{\rm e}^{-\frac{(x-x_0)}{\sqrt{D_{r,\mu}({\rm ln}t)^{r-1}}}}\;,\nonumber
\end{eqnarray}
coefficients $\kappa_0$ and $\kappa_G$ are defined by Eq. (\ref{eq74a}) for a both--sided partially permeable membrane, and are defined by the equations presented in Sec. \ref{sec5} just below Eq. (\ref{eq74a}) for a one--sidedly fully permeable membrane. 

The following question arises: whether the generalized method of images provides functions for the slow subdiffusion that is consistent with the functions obtained by means of the random walk model presented in this paper? To have the answer to this question, which will appear to be positive, we perform the following consideration.
From Eqs. (\ref{eq5}), (\ref{eq75a}) and the Laplace transform of Eq. (\ref{eq75}), which, due to Tauberian theorem, for small values of $s$ reads \cite{dk}
\begin{equation}\label{eq78a}
\hat{\omega}(s)=1-\left(\frac{{\rm ln}\mu}{{\rm ln} (1/s)}\right)^{r-1}\;, 
\end{equation}
we obtain over the limit of small values of $\epsilon$ and $s$ 
\begin{equation}\label{eq78}
\eta(\hat{\omega}(s))=1-\epsilon\frac{1}{\sqrt{D_{r,\mu}({\rm ln}(1/s))^{r-1}}}\;.
\end{equation}
From Eqs. (\ref{eq28})--(\ref{eq30a}), (\ref{eq65})--(\ref{eq67}), (\ref{eq78a}) and (\ref{eq78}), after simple calculation, we obtain functions $K_{1,1N,2,2N}$ (for the model of the reflecting membrane) and $\Lambda_{1,2}$ (for the model of the stopping membrane) for the `slow subdiffusion' case, which takes the form of the analogical functions presented in Sec. \ref{sec3} and Sec. \ref{sec4} after substituting $\sqrt{\frac{s^\alpha}{D_\alpha}}\longrightarrow \frac{1}{\sqrt{D_{r,\mu}({\rm ln}(1/s))^{r-1}}}$. Further calculation, performed using Eqs. (\ref{eq8}), (\ref{eq11}), (\ref{eq12}), (\ref{eq75a}), (\ref{eq78a}), (\ref{eq78}), and functions $K_{1,1N,2,2N}$ and $\Lambda_{1,2}$ described above, provides the Laplace transforms of Eqs. (\ref{eq77a}) and (\ref{eq77b}).

\begin{figure}[!ht]
\centering
\includegraphics[height=8.0cm]{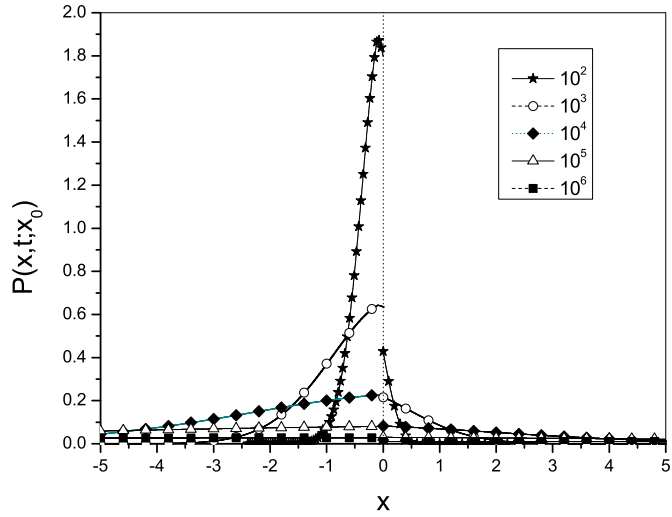}
\caption{Plots of the Green's functions for subdiffusion occurring in the system with a thin membrane (\ref{eq41}) and (\ref{eq45}), obtained for $\alpha=0.9$, $D_\alpha=0.001$, $\gamma_1=0.8$, $\gamma_2=0.3$, $x_0=-0.5$ for times given in the legend, the thin membrane, located at $x_N=0$, is represented by the dotted vertical line, all quantities are given in arbitrary chosen units.}\label{Fig3}
\end{figure}

\begin{figure}[!ht]
\centering
\includegraphics[height=8.0cm]{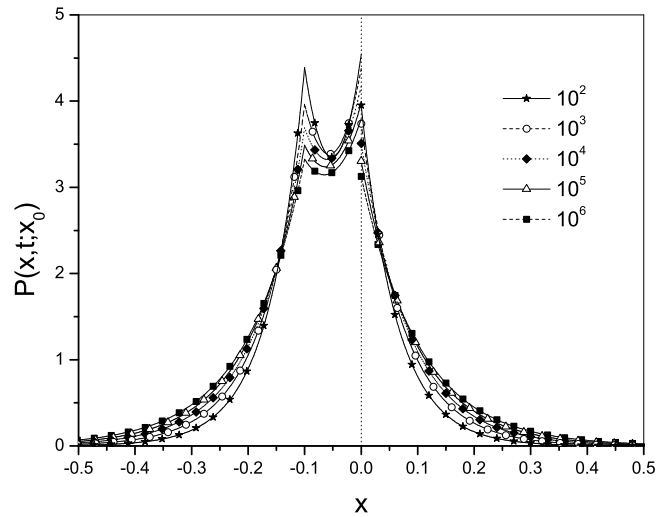}
\caption{Plots of the Green's functions for `slow subdiffusion' occurring in the system with a thin membrane (\ref{eq77a}) and (\ref{eq77b}), here $r=1.9$, $D_{r,\mu}=0.001$, the other parameters are the same as in Fig. \ref{Fig3}.}\label{Fig4}
\end{figure}

Figures \ref{Fig3} and \ref{Fig4} are merely on illustration of the Green's functions but show some general regularities of subdiffusion in a membrane system. Membrane permeability property is less manifest in a system in which slow subdiffusion occurs.
This is due to the fact that when a particle tries to jump through a thin membrane, the probability of a particle being stopped by the membrane is the same as for the different processes occurring in the membrane system but a subsequent attempt at a jump is expected to be much later in the system in which slow subdiffusion occurs. So, the probability of a particle's passing through the membrane over a certain time interval is less for this process compared to the `standard' subdiffusion process. The qualitative comparison of the changes of the Green's functions for different orders of the time variable is also interested. For `slow subdiffusion' the changes are relatively small (see Fig. \ref{Fig4}), whereas for `standard' subdiffusion the scale of the changes is much higher (see Fig. \ref{Fig3}).

\section{Final remarks \label{sec6}}

In this study we consider subdiffusion in two systems with a thin membrane. The differences between the systems are related to the assumption concerning the process of the reflecting or stopping of a particle by the membrane when an attempt to pass the particle through the membrane fails. In the first model, the particle can be reflected off by the membrane with a certain probability, the second model assumes that the particle can be stopped, with a certain probability, at the membrane.
These assumptions can be associated with the following physical interpretation. If there is a slight repulsive force, generated by the membrane and acting on the particle, then the model of the reflecting membrane can be used. Otherwise, including the case in which a small attraction of the particle exerted by the membrane is assumed, the stopping membrane model can be used to describe subdiffusion in a membrane system. 

The differences between the models are manifested in two cases:
\begin{enumerate}
	\item The probability of finding a particle at the surface of the membrane is different for both models when the membrane is partially permeable with respect to both its sides, i.e. when $q_1,q_2>0$. The Green's functions are discontinuous at the membrane surface for the model with a reflecting membrane (see Eq. (\ref{eq47})) and continuous for the model with a stopping membrane. 
	\item For the system in which the membrane is fully one--sidedly permeable, the Green's functions for both models take a form which can be expressed by the function obtained within the generalized method of images (\ref{eq71})--(\ref{eq73}) if the `gradient source function' $P_G$ is omitted. However, the Green's functions obtained for the model with a reflecting membrane and the ones obtained for the model with a stopping membrane are not equivalent to each other. The reason being that if is not possible to find the relation between the parameters $\gamma_2$ and $\tilde{\gamma}_2$ (for $q_1=0$) in such a way that Eqs. (\ref{eq50}) and (\ref{eq53}) coincide with Eqs. (\ref{eq69a}) and (\ref{eq69b}), respectively. A similar remark can be made for the case $q_2=0$.
\end{enumerate}

For the case $q_1,q_2>0$, the Green's function for both models, Eqs. (\ref{eq41}) and (\ref{eq45}) for the model with a reflecting membrane and Eqs. (\ref{eq67a}) and (\ref{eq67b}) for the model with a stopping membrane, coincide with each other, respectively, in region $(-\infty,x_N)\cup (x_N,\infty)$ if $\gamma_1=\tilde{\gamma}_1 /2$ and $\gamma_2=\tilde{\gamma}_2 /2$. Since the membrane permeability coefficients are expected to be determined from experimental data when the model is used to describe subdiffusion in a real membrane system, it does not matter which of the models will be used in the modelling unless the probability of finding diffusing particles at the membrane surfaces is not considered. 

To study the subdiffusion in a membrane system we use the model with a discrete time and space variable, next we transform the Green's functions to continuous variables by means of the formulas presented in this paper. Such a model seems to be oversimplified. However, for the homogeneous system without a membrane, it provides results which can be derived by means of more `realistic' models of subdiffusion. Thus, we assume that the model used in this paper provides the Green's functions which are useful in the modelling of subdiffusion in a system with a thin membrane.

The method of images appears to be useful tool for determining the Green's functions in a membrane system for a various model of subdiffusion for which the waiting time probability density is given by Eq. (\ref{eq75}).
It was shown that the generalized method of images allows us to obtain the Green's function for the membrane system for a `slow subdiffusion' process.
We hypothesize that the generalized method of images, presented in this paper, is applicable to other subdiffusion models. 

The considerations presented in this paper concern the case of $x_0<x_N$. In general, the coefficient $\gamma_1$ can be defined as a coefficient which controls the membrane permeability when a particle tries to pass through the membrane from a region of the initial particle's location to the opposite region, and $\gamma_2$ is a membrane permeability coefficient when the particle moves in the opposite direction. In \cite{tk} it was shown that for the case of $q_1,q_2>0$ the Green's function for the model with a stopping membrane, Eqs. (\ref{eq67a}) and (\ref{eq67b}), fulfil the following boundary condition at the membrane
\begin{equation}\label{eq79}
P_{--}(x_N^-,t;x_0)=\lambda_1 P_{+-}(x_N^+,t;x_0)+\lambda_2\frac{\partial^{\alpha/2}P_{+-}(x_N^+,t;x_0)}{\partial t^{\alpha/2}}\;,
\end{equation}
where $\lambda_1=\frac{\tilde{\gamma}_1}{\tilde{\gamma}_2}$ and $\lambda_2=\frac{\tilde{\gamma}_1}{\sqrt{D_\alpha}}$. This boundary condition is complemented by the condition of flux continuity at the membrane. The considerations presented in this paper show that boundary condition Eq. (\ref{eq79}) is fulfilled by the Green's functions at the reflecting membrane for $q_1,q_2>0$ if we assume $\tilde{\gamma}_1=2\gamma_1$ and $\tilde{\gamma}_2=2\gamma_2$. The second term in the right-hand side of Eq. (\ref{eq79}), which vanishes over a sufficiently long time (see the discussion presented in \cite{tk}), represents the `additional' memory effect created by the membrane (despite the fact that subdiffusion is a long--memory process itself). As mentioned earlier, for the case of $q_1=0$ or $q_2=0$, the Green's functions take the form of Eqs. (\ref{eq71})--(\ref{eq73}) with $\kappa_G=0$ for both models. It is easy to see that these functions fulfil the boundary condition (\ref{eq79}) in which $\lambda_2\equiv 0$. Thus, the `additional memory effect' is not created by the one--sidedly fully permeable membrane.

For the case of $x_0>x_N$, the functions and boundary conditions can be obtained from the functions presented in this paper when, due to symmetry arguments, the following conversion is made: $\gamma_1\longrightarrow \gamma_2$, $\gamma_2\longrightarrow \gamma_1$ (or $\tilde{\gamma}_1\longrightarrow \tilde{\gamma}_2$, $\tilde{\gamma}_2\longrightarrow \tilde{\gamma}_1$), $x-x_0\longrightarrow x_0-x$, $x-x_N\longrightarrow x_N-x$ and $x_N-x_0\longrightarrow x_0-x_N$. 

\section*{Acknowledgments}
This paper was partially supported by the Polish National Science Centre under grant No. 2014/13/D/ST2/03608.
The author would like to thank Dr. Katarzyna D. Lewandowska for helpful discussions.

\section*{Appendix}

Using Eqs. (\ref{eq2}), (\ref{eq3}), and (\ref{eq19})--(\ref{eq23}) we obtain
\begin{eqnarray}
 \label{a1}S(m,z;m_0)-\delta_{m,m_0}&=\frac{z}{2}S(m-1,z;m_0)+\frac{z}{2}S(m+1,z;m_0),\\
   & m\neq N-1, N, N+1, N+2,\nonumber\\
      \nonumber\\
 \label{a2}S(N-1,z;m_0)-\delta_{N-1,m_0}&=\frac{z}{2}S(N-2,z;m_0)+\frac{z(1+q_1)}{2}S(N,z;m_0),\\ 
      \nonumber\\
 \label{a3}S(N,z;m_0)-\delta_{N,m_0}&=\frac{z}{2}S(N-1,z;m_0)\\
 &+\frac{z(1-q_2)}{2}S(N+1,z;m_0),\nonumber\\ 
      \nonumber\\
 \label{a4}S(N+1,z;m_0)-\delta_{N+1,m_0}&=\frac{z(1-q_1)}{2}S(N,z;m_0)+\frac{z}{2}S(N+2,z;m_0),\\ 
      \nonumber\\
 \label{a5}S(N+2,z;m_0)-\delta_{N+2,m_0}&=\frac{z(1+q_2)}{2}S(N+1,z;m_0)\\ 
 &+\frac{z}{2}S(N+3,z;m_0).\nonumber
\end{eqnarray}

To solve Eqs. (\ref{a1})--(\ref{a5}) we use the following generating function with respect to the space variable
\begin{equation}\label{a6}
G(u,z;m_0)=\sum_{m=-\infty}^\infty u^m S(m,z;m_0)\;.
\end{equation}
The generating function $S$ can be obtained by means of the following formula
\begin{equation}\label{a7}
S(m,z;m_0)=\frac{1}{2\pi i}\oint_{K({\bf 0},1)}\frac{G(u,z;m_0)}{u^{m+1}}du\;,
\end{equation}
where integration is carried out along the unit circle $K$ centered at point ${\bf 0}=(0,0)$ in order to be consistent with an increasing argument of a complex number. From Eqs. (\ref{a1})--(\ref{a6}) we obtain
\begin{eqnarray}\label{a8}
G(u,z;m_0)=\frac{u^{m_0}}{\left[1-\frac{z}{2}\left(u+\frac{1}{u}\right)\right]}+S(N,z;m_0)\frac{zq_1}{2}\frac{\big(u^{N-1}-u^{N+1}\big)}{\left[1-\frac{z}{2}\left(u+\frac{1}{u}\right)\right]}\nonumber\\
-S(N+1,z;m_0)\frac{zq_2}{2}\frac{\big(u^{N}-u^{N+2}\big)}{\left[1-\frac{z}{2}\left(u+\frac{1}{u}\right)\right]}\;.
\end{eqnarray}
Using the integral formula
\begin{equation}\label{a9}
\frac{1}{2\pi i}\oint_{K({\bf 0},1)}\frac{u^{m_0}}{u^{m+1}\left[1-\frac{z}{2}\left(u+\frac{1}{u}\right)\right]}du=\frac{\eta^{|m-m_0|}(z)}{\sqrt{1-z^2}}\;,
\end{equation}
from Eqs. (\ref{a7}) and (\ref{a8}) we obtain, after simple calculations, Eqs. (\ref{eq24})--(\ref{eq29}).\\

\end{document}